\documentclass[amsmath,amssymb,prc,showpacs,showkeys,byrevtex]{revtex4}
\DeclareMathOperator{\Tr}{Tr}%
\DeclareMathOperator{\e}{e}
\begin{document}
\title{Probability distributions in statistical ensembles with conserved charges}
\author{J. Cleymans}
\email{cleymans@qgp.phy.uct.ac.za}
\affiliation{UCT-CERN Research
Centre and Department of Physics,University of Cape Town,
Rondebosch 7701, South Africa}
\author{K. Redlich}
\email{redlich@ift.uni.wroc.pl}
\author{L. Turko}
\email{turko@ift.uni.wroc.pl} \affiliation{Institute of Theoretical Physics, University
of Wroc{\l}aw, Pl. Maksa Borna 9, 50-204  Wroc{\l}aw, Poland}
\date{December 21, 2004}
\begin{abstract}
The probability distributions for charged particle numbers and their densities are
derived in  statistical ensembles with conservation laws. It is shown that if this limit
is properly taken then the canonical and grand canonical ensembles are equivalent. This
equivalence is proven  on the most general, probability distribution level.
\end{abstract}
\maketitle
\section{Introduction}
Macroscopic data of an equilibrium state are described by means of statistical
distributions of microscopic variables specific for a given ensemble. In the application
to the  description of particle production in high energy particle collisions we are
generally dealing with the grand canonical ensemble with respect to the particle number
\cite{pb,rh,rafelski}. In the ultrarelativistic situation energy conservation and
particle number are usually controlled by the temperature of the system \cite{liu}.

Applications of statistical physics concepts to multiparticle production processes
require however the implementation of internal symmetries \cite{redlich,turko} that
result in associated conservation of quantum numbers. In the grand canonical formulation
(GC) conservation of quantum numbers is implemented  on the average and is determined by
the corresponding chemical potential. On the other hand in the canonical ensemble (C)
quantum number conservation is treated exactly.

In general the thermodynamic quantities calculated in the GC and C ensembles differ. This
is particularly the case when dealing with small
systems~\cite{pb,rh,rafelski,liu,can,can1}. Since relativistic heavy ion collisions
correspond to a finite volume and to a given charge value, the canonical ensemble should
be used whenever the GC and C formalisms give different answers
\cite{rh,can1,ho,anti,str}. It is however to be expected that in the thermodynamic limit
the C and GC descriptions should provide the same answer for physical
observables~\cite{rh,ho,anti}. The thermodynamic limit is reached  in the large volume
for fixed density limit in C and for fixed average density  in the GC ensemble. Only in
this limit can one indiscriminately use the GC or C descriptions. The results presented
here do not rely on the system under consideration being of a relativistic nature and
apply equally well to relativistic and non-relativistic systems.

There is an essential difference in the volume dependence of observables in the GC and C
formulations \cite{pb,rh,rafelski,can,ho}. Consequently  in the limit when $V\to\infty$
some ratios of extensive quantities could in general converge to different values in GC
and C ensemble \cite{begun}. It is thus clear that the equivalence of both descriptions
in the thermodynamic limit can be strictly established only for intensive
observables~\cite{rh}.

In applications of statistical models to particle production in high energy collisions of
elementary particles \cite{pp} and in heavy ion collisions \cite{pb,aa} we are always
dealing with small systems. Thus, the model description of particle yields is in
principle canonical with respect to the conservation laws. This is particularly evident
in elementary particle and peripheral heavy ion collisions where e.g. strangeness
production is strongly suppressed due to canonical effects \cite{str}. However, a
detailed analysis of different particle yields \cite{pb,ho,anti,aa,str} has shown that in
central heavy ion collisions the relative error between the C and GC descriptions is so
small that the GC approximation can be used with confidence.

Recently, an interesting observation has been made \cite{begun} that in finite systems
even very small relative errors between  C and GC results seen on the level of first
particle moments i.e. thermal particle multiplicities do not necessarily guarantee that
this is also the case for higher moments or particle fluctuations. Contrary, for a set of
thermal parameters relevant in high energy heavy ion collisions the second  moment
differs substantially in the GC and C ensembles at finite $V$ even though the first
moments agree to high accuracy. The above observation has been generalized~\cite{begun}
to the large volume limit with the conclusion of the violation of GC and C ensemble
equivalence in the thermodynamic limit on the level of particle number fluctuations.

However, a direct  comparison of GC and C results for  the average charged particle
number or higher moments  is only adequate for a finite system. In the thermodynamic
limit such a comparison can be only done at the level of densities. This is simply
because in the large volume limit only  particle densities  are finite in GC and C
ensemble whereas corresponding  multiplicities are infinite in both cases. In the
thermodynamic limit the equivalence problem of GC and C formulation of the conservation
laws requires intensive observables.

We show that  different statistical ensembles with respect to conservation laws  are
exactly equivalent in the thermodynamic limit. Consequently, there is equivalence of all
possible moments, relative fluctuations, scaled variance etc. The exact calculations are
performed in the GC and C ensemble  with arbitrary values of the conserved charge.

\section{Probability distributions in canonical and grand canonical ensembles }
The  equivalence of canonical and grand canonical ensembles of statistical systems with
conservation laws will be discussed for a non--interacting relativistic gas of charged
particles and antiparticles  in a volume $V$ at temperature $T$. Particles and
antiparticles have charge $\pm 1$ respectively.

Let us consider a system with total overall charge $Q$ with $N=N_-$ antiparticles and
$N_+=N+Q$ particles. The requirement of the exact charge conservation in the canonical
ensemble leads to the following partition function \cite{pb,rh,rafelski}
\begin{equation}\label{eq1}
\mathcal{Z}_Q^{C}(V,T)=\Tr_Q\,\e^{-\beta\hat{H}}=
\sum\limits_{N=\text{max}(-Q,\,0)}^\infty\frac{z^{2N+Q}}{N!(N+Q)!}=I_Q(2 z),
\end{equation}
where $z$ is the one-particle partition function
\begin{equation}z(T)=\frac{V}{(2\pi)^3}\int d^3p\,\e^{-\beta\sqrt{p^2 +m^2}}=
\frac{V}{2\pi^2}Tm^2K_2\left(\frac{m}{T}\right)\equiv V z_0(T)\,.\label{eq2}
\end{equation}
$I_Q$  and $K_2$  are the modified Bessel functions \cite{Abram:2004zb}.

To obtain the average particle multiplicity  moments one introduces in  e.g.(\ref{eq1})
fugacity parameters $\lambda_+$ and $\lambda_-$, which are set to one in the final
formulae
\begin{equation}\label{eq3}
\mathcal{Z}^{C}_Q(V,T,\lambda_-,\lambda_+)=
\sum\limits_{N=\text{max}(-Q,\,0)}^\infty\frac{z^{2N+Q}}{N!(N+Q)!}\lambda_+^{N+Q}
\lambda_-^N =\left(\frac{\lambda_+}{\lambda_-}\right)^{Q/2}I_Q(2 z\sqrt{\lambda_+
\lambda_-})\,,
\end{equation}
such that
\begin{equation}
\label{eq4}
\langle N^k_\pm
\rangle_Q^C=\frac{1}{\mathcal{Z}^{C}_Q}\left.\left(\lambda_\pm\frac{\partial}
{\partial\lambda_\pm}\right)^k
\mathcal{Z}^{C}_Q(V,T,\lambda_-,\lambda_+)\right|_{\lambda_{\pm=1}}.
\end{equation}
From Eqs. (\ref{eq3}--\ref{eq4}) it is clear that
\begin{equation}\label{eq5}
    \mathcal{P}_Q^{C}(N,N+Q,V)= \frac{z^{2N+Q}}{N!(N+Q)!}\frac{1}{I_Q(2 z)}\,,
\end{equation}
is the probability distribution to find $N$ negatively and $N+Q$ positively charged
particles in a system of volume $V$, temperature $T$ and total charge $Q$ \cite{pk,lin}.

In the grand canonical ensemble the charge is conserved   on the average, thus the
partition function
\begin{equation}\label{eq7}
\mathcal{Z}^{GC}(V,T)=\Tr\e^{-\beta\hat{H} + \gamma\hat{Q}}\,,
\end{equation}
where $\gamma$ is chosen such as to reproduce the average charge $\langle Q\rangle$ in
the system. For a  non-interacting gas  the trace in the above equation can be calculated
explicitly yielding
\begin{equation}
\label{eq8} \mathcal{Z}^{GC}(V,T,\gamma,\lambda_-,\lambda_+) = \sum\limits_{N_+ =
0}^\infty \sum\limits_{N_-=0}^\infty \frac{\e^{\gamma (N_+ -
N_-)}\lambda_+^{N_+}\lambda_-^{N_-}}{N_+!N_-!}z^{N_- + N_+} =\\
\exp\left[\left(\lambda_+\e^\gamma + \lambda_-\e^{-\gamma}\right)z\right]\,
\end{equation}
where as in Eq. (\ref{eq3}) we have introduced auxiliary fugacities $\lambda_\pm$ for
particles and antiparticles. Following   Eq. (\ref{eq4}) the average number of particles
and the average charge in the GC ensemble is obtained as
\begin{equation}\label{eq9}
\langle N_\pm\rangle =z\,\exp{(\pm\gamma)}, ~~~~ \langle Q\rangle
= \langle N_+\rangle - \langle N_-\rangle =2z\,\cosh\gamma
\end{equation}
In terms of the total charge $Q=N_+-N_-$  the GC partition function (\ref{eq8}) can be
written as
\begin{equation}
\label{eq11}
\mathcal{Z}^{GC}(V,T,\gamma,\lambda_-,\lambda_+) =
\sum\limits_{N=0}^\infty\sum\limits_{Q=-N}^\infty \e^{\gamma~Q}
\frac{\lambda_+^{Q+N}\lambda_-^{N}}{N!(Q+N)!}z^{Q+2N} ,
\end{equation}
with $N=N_-$ being the number of negatively charged particles. Thus, the function
\begin{equation}
\label{eq12} \mathcal{P}_{\langle Q\rangle}^{GC}(N,N+Q,V) =
\frac{1}{\mathcal{Z}^{GC}}\frac{z^{Q+2N}}{N!(Q+N)!}\e^{\gamma Q}
\end{equation}
defines  the probability distribution  in a GC ensemble, with average charge $\langle
Q\rangle$, to find  a charge $Q$ with $N$ negatively charged particles. Expressing the
chemical potential appearing in (\ref{eq12}) through the corresponding average charge
from (\ref{eq9}) one finds
\begin{equation}
\label{eq13}
\mathcal{P}_{\langle Q\rangle}^{GC}(N,N+Q,V) =
   \frac{z^{2N+Q}}{N!(N+Q)!}\left[\frac{\langle Q\rangle +
    \sqrt{\langle Q\rangle^2+4 z^2}}{2 z}\right]^Q \e^{-\sqrt{\langle Q\rangle^2+4~z^2}}\,.
\end{equation}
All other  less restricted probabilities can now be obtained directly from Eq.
(\ref{eq13}). The probability distribution $\mathcal{P}^{GC}_{\langle Q\rangle}~(N,V)$ to
find $N$ particles or $\mathcal{P}^{GC}_{\langle Q\rangle} (Q,V)$ to find charge $Q$ in
the volume $V$ at a given average charge $\langle Q\rangle$ are obtained from
\begin{subequations}
  \begin{eqnarray}
   \mathcal{P}_{\langle Q\rangle}^{GC}(N,V) &=&
   \sum\limits_{Q=-N}^\infty\mathcal{P}_{\langle Q\rangle}^{GC}(N,N+Q,V)\,,
   \label{eq14a}\\
  \mathcal{P}_{\langle Q\rangle}^{GC}(Q,V)&=&
  \sum\limits_{N=\text{max}(-Q,\,0)}^\infty\mathcal{P}_{\langle Q\rangle}^{GC}(N,N+Q,V)\,,
\label{eq14b}
 \end{eqnarray}
\end{subequations}
The summations in Eqs. (\ref{eq14a}) and (\ref{eq14b}) can be done explicitly yielding
\begin{subequations}
\begin{equation}\label{eq15a}
\mathcal{P}_{\langle Q\rangle}^{GC}(N_\pm,V) =
   \frac{1}{N_\pm!}\left[\frac{\sqrt{\langle Q\rangle^2+4 z^2}\pm \langle Q\rangle}{2}\right]^{N_\pm}\,
   \exp\left[-\frac{\sqrt{\langle Q\rangle^2+4 z^2}\pm \langle Q\rangle}{2}\right]\,.
\end{equation}
For the charge distribution one finds
\begin{equation}\label{eq15b}
\mathcal{P}_{\langle Q\rangle}^{GC}(Q,V)= I_Q(2z)\left[\frac{\langle Q\rangle + \sqrt{\langle Q\rangle^2+4 z^2}}{2 z}\right]^Q \e^{-\sqrt{\langle Q\rangle^2+4  z^2}}\,.
\end{equation}
\end{subequations}
The particle number  probability distribution (\ref{eq15a}) is,
 as  expected~\cite{pk,lin}, a Poisson distribution.
The charge distribution, on the other hand, is not Poissonian due to the constraints
imposed by the requirement of the exact charge conservation in a given sector of GC
ensemble with a fixed average charge $\langle Q\rangle$.
\section{Probability distributions in the thermodynamic limit}
In the previous section we have introduced a set of  probability distributions in the GC
and C ensembles related to the charged particle number. All these distributions are also
valid for any value of the conserved charge as well as  the volume of the system. Thus,
they can be used to analyze the thermodynamic limit.

From~(\ref{eq5}) and (\ref{eq13}) one finds
\begin{equation}
\label{eq16} \mathcal{P}_{\langle Q\rangle}^{GC}(N,N+Q,V) =
\mathcal{P}_Q^{C}(N,N+Q,V)\,\mathcal{P}_{\langle Q\rangle}^{GC}(Q,V)\,.
\end{equation}
It is straightforward that in the sector of fixed  charge $Q$ the corresponding particle
number is distributed as   in the  canonical ensemble. That is why in Eq. (\ref{eq16})
the GC probability function $\mathcal{P}_{\langle Q\rangle}^{GC}$ is just the product of
the canonical particle number and the grand canonical charge distribution.

To take  the thermodynamic limit in  (\ref{eq16}) one first  expresses the variables
$(N,Q,\langle Q\rangle )$ by means of the corresponding densities $(n,q,\langle  q
\rangle )$ and then one takes the limit $V\to \infty$ for fixed densities. This also
requires   the replacement of a discrete sum  $(1/V)\sum_N \to \int dn$.

The essential difference between GC and C distributions in Eq.
(\ref{eq16}) appears through the probability function
$\mathcal{P}_{\langle Q\rangle}^{GC}(Q,V)$. Thus, to study the
equivalence of GC and C ensemble on the probability level in the
thermodynamic limit, it is sufficient to find $\mathcal
{P}_{\langle  q\rangle}^{GC^\infty}( q )$ from
\begin{equation}\label{eq17}
\mathcal {P}_{\langle  q\rangle}^{GC^\infty}( q ) = \lim_{V\to\infty}V
\mathcal{P}_{\langle  q\rangle}^{GC}( q ,V),
\end{equation}
where
$\mathcal {P}_{\langle  q\rangle}^{GC}$
is obtained from Eq. (\ref{eq15b})
by replacing ${\langle Q\rangle}$ and $Q$ by ${V\langle  q\rangle}$ and $V q$ respectively.
The large volume limit in (\ref{eq17}) is taken at fixed densities. An extra volume
factor in Eq. (\ref{eq17}) appears from the replacement of discrete by continuum
variables.

The limit in Eq. (\ref{eq17}) is obtained from the $\alpha\to\infty$ behavior of the
Bessel function \cite{Abram:2004zb}
\begin{equation}
\label{eq18} I_\alpha(\alpha
x)\simeq~\frac{\e^{\alpha\sqrt{1+x^2}}}{\sqrt{2\pi\alpha}(1+x^2)^{1/4}}\left[\frac{x}{1+\sqrt{1+x^2}}\right]^\alpha
\end{equation}
Consequently, the charge density probability distribution
\begin{equation}
\label{eq19}
  {\mathcal {P}}^{GC^\infty}_{\langle q\rangle}( q)
=\lim_{V\to\infty}\left\{V^{1/2} \frac{\e^{-V(~ \sqrt{\langle q\rangle^2+4z_0^2}-\sqrt{
q^2+4z_0^2}~) }}{\sqrt{2\pi}( q^2+4z_0^2)^{1/4}}\left[\frac{\langle
q\rangle+\sqrt{\langle q\rangle^2+4z_0^2}}{ q+\sqrt{ q^2+4z_0^2}}\right]^{V q}\right\}\,.
\end{equation}
It is rather straightforward to see from Eq. (\ref{eq19}) that the limit $V\to\infty$
does not exist as a regular function. This limit is zero for any $ q\neq\langle q\rangle$
and infinity for $ q =\langle q\rangle$. Let us consider however the thermodynamic limit
in Eqs. (\ref{eq17}) and (\ref{eq19}) as a generalized function
\begin{equation}\label{eq20}
\mathcal{F}=\lim_{V\to\infty}V\int dq {P}_{\langle  q\rangle}^{GC}( q ,V)f( q )\,.
\end{equation}
The density integration in (\ref{eq20}) is obtained through the saddle-point method
\begin{equation}
\label{eq21}
\mathcal{F}=\lim_{V\to\infty}V^{1/2}\int dq~e^{VS( q )}~f(q),
\end{equation}
where the function
\begin{equation}
\label{eq22} S(q)=\sqrt{ q^2+4z_0^2} - \sqrt{\langle q\rangle^2+4z_0^2} +
q\log\left(\langle q\rangle+\sqrt{\langle q\rangle^2+4z_0^2}\right) - q\log\left( q +
\sqrt{ q^2+4z_0^2}\right)\,.
\end{equation}
In the large volume limit  the dominant contribution to the integral  (\ref{eq21}) is
obtained as
\begin{equation}
\label{eq23}
\mathcal{F}=\lim_{V\to\infty}V^{1/2}
\left\{\sqrt{-\frac{2\pi}{V S^{''}( q_0)}}\,g(~q_0)\e^{V S( q_0)}+\mathcal{O}(V^{-3/2})\right\},
\end{equation}
where $q_0=\langle  q\rangle$ is just a saddle--point such that $S^{'}( q_0)=0$. From
Eq. (\ref{eq23}) one finds
\[\mathcal{F}=f(\langle  q\rangle)\,,\]
which means that the charge density probability distribution (\ref{eq17}) converges to a
delta function
\begin{equation}
\label{eq24}
\mathcal {P}_{\langle  q\rangle}^{GC^\infty}( q )=\delta ( q-\langle  q\rangle ).
\end{equation}
The above result together with Eq. (\ref{eq16}) taken in the thermodynamic limit
completes the proof of equivalence of the GC and C ensembles on the probability level.
The probability to find a given density of particles and antiparticles in the GC ensemble
with a fixed average charge density $\langle  q\rangle$ is exactly equal to the
corresponding probability in the C ensemble if one identifies  charge density $ q$ in the
C ensemble with $\langle  q\rangle$. The same  replacement $ q\to \langle  q\rangle$ is
required for all intensive thermodynamic observables.

\section{Summary}
The equivalence of the grand canonical and canonical descriptions of the conservation
laws has been considered in  the thermodynamic limit. The problem has been studied in the
relativistic gas composed only from one type of non--interacting particles and
antiparticles with the constraints imposed  by  charge conservation.  Detailed studies of
the equivalence problem of canonical  and grand canonical ensemble were presented on the
level of  probability distributions. In the thermodynamic limit the probability
distributions for particle and antiparticle densities coincide in canonical and grand
canonical ensembles. Consequently, for charged particle densities there is full
equivalence of all possible scaled moments, relative fluctuations and scaled variance in
the thermodynamic limit.
\begin{acknowledgments}
  We acknowledge the stimulating discussions with A. Ker\"anen. K.R.~also acknowledges
fruitful discussions with R. Gavai and H. Satz. This work is
partially  supported by the Polish Committee for Scientific
Research under contract KBN~2~P03B~069~25 and the Polish--South
African Science and Technology cooperation project.
\end{acknowledgments}

\end{document}